\documentclass[12pt]{article} 
 
\usepackage{epsf,epsfig} 
\usepackage{graphics} 
\usepackage{bbold} 
\usepackage{dsfont} 
\usepackage{xcolor}  
\usepackage{cite} 
\usepackage{slashed} 
\usepackage{verbatim} 
\usepackage{amsmath,amssymb,booktabs} 
\usepackage{float} 
\def\be{\begin{equation}} 
\def\ee{\end{equation}} 
\def\bea{\begin{eqnarray}} 
\def\eea{\end{eqnarray}}

\def\nn{\nonumber}

\begin{document} 
 
\allowdisplaybreaks 
\thispagestyle{empty}   
   
\begin{flushright}   
{\small   
TUM-HEP-1012/15\\ 
OUTP-15-21P\\  
\today    
}   
\end{flushright}   
\vskip1.5cm   
 
\begin{center} 
\Large\bf\boldmath 
$\bar B\to X_s \gamma$ with a warped bulk Higgs 
\unboldmath 
\end{center} 
 
\vspace{1cm} 
\begin{center} 
{\sc P.~Moch$^{a}$} and  
{\sc J.~Rohrwild$^{b}$}\\[5mm] 
{\sl ${}^a$Physik Department T31, James Franck-Stra\ss e 1\\ 
Technische Universit\"at M\"unchen,\\ 
D--85748 Garching, Germany}\\ 
\vspace{0.3cm} 
{\sl ${}^b$Rudolf Peierls Centre for Theoretical Physics,\\ University of Oxford, 
1 Keble Road,\\ Oxford OX1 3NP, United Kingdom}\\[1.5cm] 
\end{center} 
 
\date{\today} 
 
\begin{abstract} 
  We study the decay $\bar B\to X_s\gamma$ in Randall-Sundrum models
  with an IR-localised bulk Higgs. The two models under consideration are
  a minimal model as well as a model with a custodial protection mechanism.
  We include the effects of tree- and one-loop diagrams involving 5D gluon 
  and Higgs exchanges as well as QCD corrections arising from the evolution 
  from the Kaluza-Klein scale  to the typical scale of the decay. 
  We find the RS corrections to the branching fraction can be sizeable for large
  Yukawas and moderate KK scales $T$; for small Yukawas the RS contribution is
  small enough to be invisible in current experimental data.
 
 \end{abstract} 
 
\newpage   
\setcounter{page}{1}

\newpage 

\section{Introduction} 
\label{sec:intro} 
One of the best studied processes in flavour physics is the 
inclusive radiative $\bar B\to X_s \gamma$ decay. On the experimental 
side numerous experiments \cite{Saito:2014das,Lees:2012ym,Lees:2012ufa,Lees:2012wg,
Limosani:2009qg,Aubert:2007my,Chen:2001fja,Abe:2001hk} 
provide an ever increasing amount of data; leading to the current 
HFAG average \cite{Amhis:2014hma} of 
\begin{align}
\label{eq:bsgammaExp}
 {{\rm{Br}}(\bar B\to X_s \gamma)^{exp}_{E_\gamma>1.6{\rm GeV}}}= (342\pm 21 \pm 7)\times 10^{-6}\,,
\end{align}
where all contributing experimental results were converted as 
to correspond to a lower photon energy cut of $1.6\,{\rm GeV}$. 
A further improvement of this number can be anticipated:
the Belle II experiment is expected to be able to measure the branching fraction 
with an uncertainty of about $6\%$ \cite{Aushev:2010bq}. 

On the theory side, the fact that the rare radiative decay provides both  
powerful check for the Standard Model (SM) of particle physics
and is sensitive to physics beyond the SM (BSM) fuelled a tremendous effort (see 
e.g.~\cite{Misiak:2006zs,Misiak:2006ab,Czakon:2006ss,Czakon:2015exa} and 
references therein) to understand the intricacies of the $b\to s\gamma$ transition. 
The most recent result \cite{Misiak:2015xwa} is given by
\begin{align}
\label{eq:bsgammaSM}
  {{\rm{Br}}(\bar B\to X_s \gamma)^{th}_{E_\gamma>1.6{\rm GeV}}}= (336\pm 23)\times 10^{-6}\;.
\end{align}
It is in very good agreement with experiment, cp.~\eqref{eq:bsgammaExp}, and 
therefore provides non-trivial constraints to any New Physics model that can 
generate additional flavour-changing neutral currents (FCNCs). 

Extra-dimensional models of the Randall-Sundrum (RS) type \cite{Randall:1999vf} are known to have 
a particularly rich flavour phenomenology and can, despite 
an inherent protection mechanism \cite{Agashe:2004cp}, give rise to sizeable FCNCs.
The characteristic five-dimensional metric of RS models 
can be written as
\begin{equation}
ds^2 = \left(\frac{1}{k z}\right)^{\!2}\left(\eta_{\mu\nu} dx^\mu dx^\nu-
dz^2\right), 
\label{metric}
\end{equation}
in conformal coordinates. Here $k = 2.44 \cdot 10^{18}\,$GeV is of order of the Planck 
scale $M_{\rm Pl}$. The fifth coordinate $z$ is restricted to 
the interval $[1/k,1/T]$. The boundaries  $z=1/k$ and $z=1/T$ are typically referred 
to as Planck and IR brane respectively. The a priori arbitrary scale $T$ is assumed
of the order of a TeV in order to alleviate gauge-gravity hierarchy issues 
\cite{Randall:1999ee}.

One of the main reasons for the popularity of these models is
the interplay of (SM) flavour and properties of 5D wave functions
\cite{Gherghetta:2000qt, Huber:2000ie, Huber:2003tu}.
In particular,  mass and CKM hierarchies can be related to 
the strength of the Planck or IR brane localisation of the corresponding 
KK zero-mode wave functions \cite{Grossman:1999ra}. This intimate relationship of 
geometry and flavour makes the study of flavour physics observables 
all the more intriguing. For most processes like meson mixing 
\cite{Csaki:2008zd, Blanke:2008zb} or electroweak
pseudo-observables \cite{Casagrande:2008hr,Casagrande:2010si,Cabrer:2011vu} 
the RS contribution arises (to leading order) from 
tree-level corrections to dimension-six operators, e.g., four-quark 
operators in the case of meson mixing.

In the last few years loop-induced processes, that is processes that to leading order 
do not receive  contributions from tree-level diagrams in RS models, 
have been studied quite extensively.
Observables that have been investigated include $\mu \to e \gamma$ 
\cite{Agashe:2006iy,Csaki:2010aj,Beneke:2015lba}, $(g-2)_\mu$ 
\cite{Beneke:2012ie,Moch:2014ofa},
Higgs production and decay 
\cite{Azatov:2010pf,Carena:2012fk,Malm:2013jia,Hahn:2013nza,Archer:2014jca,Malm:2014gha} 
as well as $c\to u \gamma$ and $c\to u g$ \cite{Delaunay:2012cz}.
The latter process just as $\mu \to e \gamma$ receives contributions 
from Kaluza-Klein (KK) states of the Higgs (in models 
where these are present). The subtleties involving their determination have 
only recently been pointed out \cite{Agashe:2014jca}.

The decay $\bar B\to X_s \gamma$ has been studied previously in the context of RS models 
in \cite{Blanke:2012tv} in the 5d picture and in \cite{Biancofiore:2014wpa} using a Kaluza-Klein mode decomposition.
\cite{Biancofiore:2014wpa} maintains its focus on the decay $\bar B \to K^* \,\mu^+\,\mu^-$.  Both works consider only the dominant effects of 5D penguin
diagrams and neglect the so-called wrong-chirality Higgs couplings terms
\cite{Agashe:2006iy, Azatov:2009na}. This is equivalent to an RS model with a 
naively brane-localised Higgs that does not arise from a well-defined
limiting procedure.

In this letter we want to consider the case of a bulk Higgs field.
This scenario is quite general as it requires us to take into account 
both Higgs and KK Higgs contributions. In order to keep the advantages of 
the original setup, we still impose that the bulk Higgs is strongly IR localised. 
Following the construction of \cite{Cacciapaglia:2006mz} for the bulk Higgs
gives a 5D wave function for the Higgs vacuum expectation value (vev) of
the form
\begin{align}
  v(z)=\sqrt{\frac{2(1+\beta)}{1-\epsilon^{2+2\beta}} } \,
  k^{3/2} T^{\beta+1} v_{\rm SM} \,z^{\beta+2}. 
\label{eq:vevprofile}
\end{align}
Here $\epsilon=T/k$, $v_{\rm SM}\equiv v\simeq 246\,\mbox{GeV}$ and $\beta$ is a parameter
related to the 5D mass of the Higgs scalar. The typical width of the zero-mode profile is 
determined by $1/(\beta T)$; the limit $\beta\to \infty$ leads to a maximally 
localised 'bulk' Higgs. For our subsequent analysis we always tacitly assume that this 
limit has been taken.\footnote{See \cite{Beneke:2015lba} for details on how the 
limit has to be taken if 5D loops with a (dimensional) regulator are involved.}
We will focus on two types of RS models: the minimal model with the same gauge and fermion 
multiplets as the Standard Model and the custodially protected model \cite{Agashe:2003zs,Agashe:2006at}
with an extended matter and gauge sector (see \cite{Albrecht:2009xr} for details on the specific setup).

This setup was also used in our work of lepton flavour violation \cite{Beneke:2015lba}
and we refer the reader to it and to \cite{Beneke:2012ie} for explicit expressions for
the 5D action and associated Feynman rules.
For the study of the $b\to s \gamma$ transition we can directly transfer the results of \cite{Beneke:2015lba}
to the quark sector. For simplicity, we only consider the effects of the strong interaction and 
the Higgs boson. Electroweak effects could be included in full analogy to the 
existing computation of flavour violation in the lepton sector, however, their
inclusion will not lead to a fundamentally different phenomenology.
Since we focus on QCD effects, we do not investigate the phenomenologically interesting 
decay $B\to K^\ast \ell^+\ell^-$; it receives tree-level contributions from four-fermion operators with both 
quark and lepton fields, which cannot be generated by gluon exchanges.

The general strategy of the calculation then follows \cite{Beneke:2012ie}.
We start with a fully 5D theory and integrate out the compact fifth dimension by  
matching onto an effective Lagrangian at the KK scale $T$. We will only consider operators
of at most dimension six and the corresponding effective Lagrangian is the renowned Buchm\"uller-Wyler 
Lagrangian \cite{Buchmuller:1985jz}. This step is presented in section \ref{sec:Initial}.

We then transition from the symmetric phase to the broken electroweak phase. The Wilson
coefficients of the resulting operators are subsequently evolved from the high scale $T$ down to the 
typical scale of the process $b\to s\gamma$, $\mu_b$. This is discussed in Sec.\ref{sec:running}. 
The phenomenological implications of the resulting corrections to the coefficients in the 
weak Hamiltonian are shown in section \ref{sec:pheno}. We conclude in Sec.\ref{sec:conclusion}.

\section{Matching at the scale $T$} 
\label{sec:Initial} 

A starting point for a completely general analysis of 
flavour-violating processes in BSM models is the Buchm\"uller-Wyler 
Lagrangian \cite{Buchmuller:1985jz}. The new 
heavy degrees of freedom have been removed by matching onto
the effective (dimension-six) Lagrangian. This will capture
the dominant effects of any new physics model and only SM fields
and dynamics are needed in any subsequent analysis. The price for 
taming a BSM model in this way is encoded in the (potentially) up to $2499$ Wilson 
coefficients{\footnote{If all possible flavour structure are counted\cite{Jenkins:2013zja}.}} 
Each of these has to be determined by integrating out heavy degrees of freedom 
above the matching scale.

Here we are only interested in the dominant contribution 
to $b \to s$ transitions in a specific class of RS models.
That is, we only consider the flavour-changing transitions 
that are mediated by KK gluons and the (KK) Higgs. 
This greatly limits the number of operators that have to be 
considered.
It is then convenient to consider the following    
effective Lagrangian at the KK scale $\mu_{KK}=T$.
\begin{align}
\label{eq:LagraniganHigh}
\mathcal{L}^{dim\,6}\supset \frac{1}{T^2}\Big[& \;  a_{ij}^{g} \bar Q_i \Phi \sigma^{\mu\nu} T^A D_j \,G^A_{\mu\nu} 
                          + a_{ij}^{\gamma} \bar Q_i \Phi \sigma^{\mu\nu}  D_j\, F_{\mu\nu}  + \text{h.c.} \nn \\
		&	  + b^{QQ}_{ij}  \bar Q_i \gamma^\mu T^A Q_i \; \bar Q_j \gamma_\mu T^A Q_j
			  + b^{QU}_{ij}  \bar Q_i \gamma^\mu T^A Q_i \; \bar U_j \gamma_\mu T^A U_j \nn \\
		&	  + b^{QD}_{ij}  \bar Q_i \gamma^\mu T^A U_i \; \bar D_j \gamma_\mu T^A D_j
			  + b^{DD}_{ij}  \bar D_i \gamma^\mu T^A D_i \; \bar D_j \gamma_\mu T^A D_j \nn \\
		&	  + b^{DU}_{ij}  \bar D_i \gamma^\mu T^A D_i \; \bar U_j \gamma_\mu T^A U_j\nn \\
		&  +\ldots\;\Big]\;,
\end{align}
where we dropped operators that either will not contribute to leading logarithmic (LL) accuracy to 
$b\to s\gamma$ or are generated by exchange of SU$(2)$, U$(1)$ gauge bosons. 
$Q_i$ corresponds to a quark doublet of with generation index $i$; $D$ and $U$ are
down- and up-type singlets. $G$ and $F$ are gluonic and electromagnetic field strength tensor, respectively; 
$T^A$ is a generator of $SU(3)$ in the fundamental representation. 
Note that the Lagrangian is defined in the unbroken electroweak phase and all quarks are
still massless. Hence the indices $i,j=1,2,3$ are not commensurate with 
e.g.~up, charm or top.
The ellipses indicate a sizeable set of omitted operators that either cannot be generated 
via QCD effects or whose contribution to $b\to s \gamma$ is suppressed.

In  writing \eqref{eq:LagraniganHigh} we tacitly assumed 
that we are in a flavour basis where the 5D fermion mass matrix is diagonal.
Furthermore, \eqref{eq:LagraniganHigh} already reflects the fact that we will need 
the coefficient of the electromagnetic dipole operator, i.e.~instead of 
working with the field strength tensors of SU$(2)_L$ and U$(1)_Y$ we only included
the linear combination that will form the photon after EWSB. Using Fierz transformations
it is possible to rewrite some of the operators in \eqref{eq:LagraniganHigh} by removing 
the $T^a \otimes T^a$ colour structure. This procedure is useful for a general analysis
of flavour violation as one can use a minimal operator basis \cite{Grzadkowski:2010es}.
For our simplified analysis this is not needed.

The Wilson coefficients $a$ and $b$ will set the initial conditions 
for the RGE evolution from $\mu_{KK}$ to the electroweak scale $\mu_{EW} \sim M_W \sim m_{top}$
where they will induce shifts in the coefficients of the well-known weak Hamiltonian. 
A subsequent evolution down to the scale $\mu_b$ can then be performed in the standard 
way.

Before moving on to the results for the matching calculation let us
briefly review the parameters of the RS model that are relevant to 
our analysis. As for any BSM study of flavour the Yukawa matrices are
of crucial importance. An RS Lagrangian incorporates two 5D dimensionless 
Yukawa matrices, $Y_u$ and $Y_d$, corresponding to the couplings of the Higgs to
up- and down-type SU(2)$_L$ singlets. We always impose that these matrices are 
anarchic, that is, the matrix elements are roughly of $\mathcal{O}(1)$  and  
have random phases.
Furthermore, as already mentioned above, each 5D Lagrangian (independent 
of the presence of a custodial protection mechanism) contains a 5D mass 
$M_{\psi_i}$ for each 5D fermion field $\psi_i$.
In practice, it is convenient to work with dimensionless parameters $c_{\psi_i}= M_{\psi_i}/k$.
Hence, we have in total nine 5D mass parameters: $c_{Q_i},\,c_{U_i},\,c_{D_i}$ with
$i=1,2,3$.
In order to obtain a phenomenologically viable low-energy theory that 
reproduces not only the SM quark masses but also the CKM matrix 
the mass parameters cannot be completely unrelated. E.g.~$c_{D_i}$,
the mass parameters for the down-type singlets are usually not too 
far from $-0.5$. See \cite{Casagrande:2008hr} for details on the 
relation of the various parameters for anarchic RS models.

\subsection{Gluon-mediated four-fermion operators}
\label{sec:gluonic4fermion}
The simplest way to match the 5D theory in AdS$_5$ onto the 
effective Lagrangian is using 5D Feynman rules \cite{Randall:2001gb}. This method 
is well established, see e.g.~\cite{Csaki:2010aj,Beneke:2012ie,Malm:2013jia} 
for various applications.
In particular, it avoids dealing directly with KK sums at the price
of a more complicated integral structure in loops diagram.
However, for operators that can be generated by tree-level interactions 
in the 5D theory no such complications occur and 
the matching calculation is straightforward.

A further simplification for tree-level matching comes from the 
fact that there is only a very limited number of 5D vertex integrals
that can occur. In particular, for intermediate gluons there is only one 
independent structure. The four-fermion Wilson coefficients differ only 
by symmetry factors and 5D mass parameters.
We can then use the more general results of \cite{Beneke:2015lba} for the matching 
onto four-lepton operators. One only needs 
adjust the couplings and gauge-group factors accordingly.

We find e.g.~for the Wilson coefficient of the operator $\bar Q_i \gamma^\mu T^A Q_i  \bar D_j \gamma_\mu T^A D_j$
\begin{align}
 b^{QD}_{ij} &= (i g_s)\frac{\ln(k/T)}{k} T^2 \int\limits_{1/k}^{1/T}\frac{dx}{(kx)^4}\int\limits_{1/k}^{1/T}\frac{dy}{(ky)^4} {f^{(0)}_{Q_i}(x)}^2
                                 {f^{(0)}_{D_j}(x)}^2  \Delta^{ab,ZMS}_{\text{gluon}}(0,x,y)\;.
\end{align}
where the zero-mode subtracted 5D gluon propagator is given by \cite{Beneke:2012ie}
\begin{align}
 \Delta^{ab,ZMS}_{\text{gluon}}(q\to 0,x,y) = 
 \Theta(x-y)\delta^{ab}& \,\frac{i k}{\ln\frac{k}{T}} 
\,\bigg(\frac{1}{4}\,
\bigg\{\frac{1/T^2-1/k^2}{\ln\frac{k}{T}} - x^2-y^2
+ 2 x^2 \ln(x T)\nonumber\\
& \hspace*{0cm} + \, 
2 y^2 \ln(y T)+2 y^2 \ln\frac{k}{T}
\bigg\} + {\cal O}(q^2) \bigg) 
+ (x\leftrightarrow y)
\end{align}
and the 5D wave functions are
\begin{align}
 f^{(0)}_{Q_i}(x) = T^{1/2-c_{Q_i}}k^2 x^{2-c_{Q_i}}  \sqrt{\frac{1-2 c_{Q_i}}{1-\epsilon^{1-2c_{Q_i}}}} \;&&\;
 g^{(0)}_{{D/U}_i}(x) = T^{1/2+c_{{D/U}_i}} k^2 x^{2+c_{{D/U}_i}}  \sqrt{\frac{1+2 c_{{D/U}_i}}{1-\epsilon^{1+2c_{{D/U}_i}}}} 
\end{align}
where $\epsilon=T/k$.

Up to terms suppressed by the ratio $\epsilon$
one then obtains 
\begin{equation}
b^{QD}_{ij} = b_0 + b_1(c_{Q_i}) + b_1(-c_{D_j}) 
+ b_2(c_{Q_i},c_{D_j}) 
\end{equation}
with \cite{Beneke:2012ie}
\begin{eqnarray}
&& b_0 = - \frac{ {g_s}^2}{4}\,\frac{1}{\ln(1/\epsilon)} ,
\nonumber\\
&& b_1(c) = - \frac{ {g_s}^2}{4}\,\frac{(5-2 c)(1-2 c)}{(3-2
  c)^2} \,\frac{\epsilon^{2 c-1}}{1-\epsilon^{2 c-1}},
\\
 && b_2(c_Q,c_D) = 
 - \frac{ {g_s}^2}{2}\,
\frac{(1-2 c_Q) (1+2 c_D)(3-c_L+c_D)}
{(3-2 c_Q) (3+2 c_D)(2-c_L+c_D)}
 \,\ln\frac{1}{\epsilon}\,
 \frac{\epsilon^{2 c_Q-1}}{1-\epsilon^{2 c_Q-1}}
 \,\frac{\epsilon^{-2 c_D-1}}{1-\epsilon^{-2 c_D-1}}.
\nonumber
\end{eqnarray}

The Wilson coefficients of all other operators are related to 
$b^{QD}_{ij}$. They only differ by symmetry factors
that take into account the exchange of identical quarks
and the potentially different external wave functions $f^{(0)}$ and  $g^{(0)}$.
In particular, one finds
\begin{align}
  b^{QU}_{ij}&=b^{QD}_{ij}\{c_{D_j} \to c_{U_j}\} & b^{QQ}_{ij}&=\frac{1}{2}b^{QD}_{ij}\{c_{D_j} \to -c_{Q_j}\} \\
  b^{DD}_{ij}&=\frac12 b^{QD}_{ij}\{c_{Q_i} \to -c_{D_i}\} & b^{UD}_{ij}&= b^{QD}_{ij}\{c_{Q_i} \to -c_{U_i}\} \;. 
\end{align}

\subsection{Dipole operators}
\label{sec:dipole}

\begin{figure}
 \centering
 \includegraphics[width=0.8\textwidth]{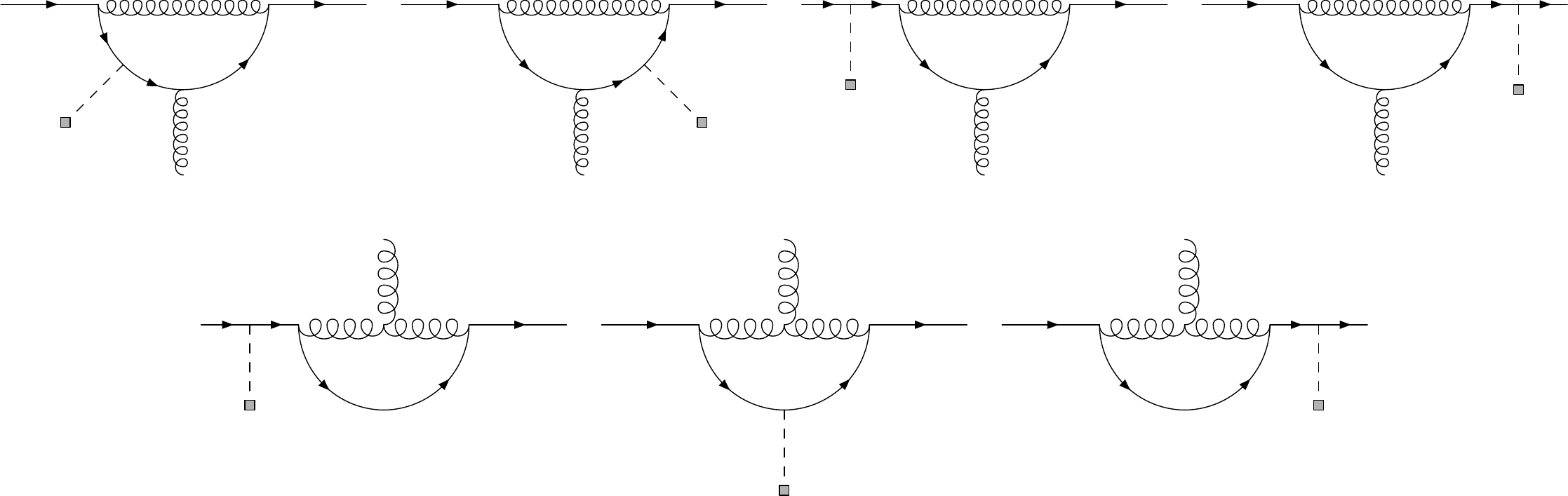}
 \caption{Topologies of 5D one-loop diagrams that contribute to the
          matching onto $a^{\gamma/g}$ at order $\alpha_s$. The external boson 
          in the diagrams in first line can be a gluon or a photon. Internal bosons 
          lines represent a 5D gluon propagator. \label{fig:topologies}}
\end{figure}
The determination of the dipole coefficients $a^\gamma$ and $a^g$
is much more involved. Following the calculation of \cite{Beneke:2012ie} the 
$\mathcal{O}(\alpha_s)$ contribution to $a^\gamma$ requires the 
computation of the diagrams shown in the upper row of figure~\ref{fig:topologies}. The contribution
to $a^g$ involves the same diagrams (with the external photon replaced by a gluon) 
and the additional non-abelian diagrams shown in the lower row of 
figure \ref{fig:topologies}. Since the determination 
of the electromagnetic dipole operators for leptons requires all  
topologies (see \cite{Beneke:2012ie, Moch:2014ofa}) both $a^\gamma$
and $a^g$ can be obtained from known results by rescaling each individual diagram
with a simple factors. This also implies directly that the 
5D $R_\xi$ gauge invariance check for the leptonic calculation  \cite{Beneke:2012ie, Moch:2014ofa}
can be carried over to the case of diagrams with (KK) gluons.

Let us consider an example: The first diagram in the first row of figure~\ref{fig:topologies}
with both the internal and the external boson gluons.
The contribution to $a^g$ can be obtained from the known result 
for same diagram topology with an external photon and 
an internal hypercharge boson $B$. Starting from this result 
we set all fermion hypercharges
$Y_f$ to $2$, trade the U(1) couplings $g^\prime$ for $g_s$  and replace the 
global factor $i  Q_f e$ from the photon vertex
with $- \frac{1}{2 N_c} i g_s T^A$. All other diagrams can be 
determined analogously.

\begin{figure}
 \centering
 \includegraphics[width=0.2\textwidth]{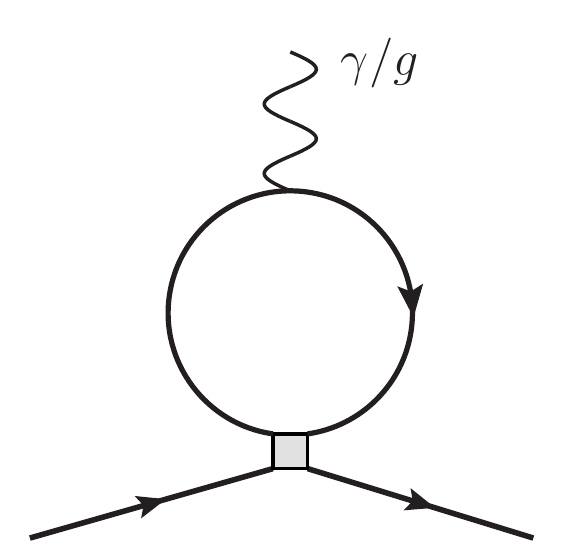}
\caption{Diagram topology that cancels the residual scheme dependence in $a^{\gamma/g}$. 
         The box denotes an insertion of a four-fermion operator.
 \label{fig:schemediag}}
\end{figure}

The way the computation of the dipole operator coefficients in \cite{Beneke:2012ie} is 
set up, we need to include contributions to the dipole structure from one-loop diagrams with an insertion 
of a four-quark operator, see Figure~\ref{fig:schemediag}. These extra terms ensure that
the Wilson coefficient is scheme independent. This otherwise 
occurring scheme dependence is a well-known fact in flavour physics, see
e.g.~\cite{Ciuchini:1993ks,Ciuchini:1993fk}.
By adding the contribution of the four-quark operators we can work with a scheme 
independent ``effective dipole coefficient'' analogous to 
the construction of \cite{Buras:1993xp}. 

Due to the required chiral structure only four-quark operators that involve both 
doublet $Q$ and singlets $D$ can contribute: $\bar Q_i \gamma^\mu T^A U_i  \bar D_j \gamma_\mu T^A D_j$. 
Up to a trivial colour factor this additional contribution is then 
again completely analogous to the one in the lepton case and we 
refer to \cite{Beneke:2012ie} for a detailed calculation.

\subsubsection{Higgs contributions}
\label{sec:Higgsterms}

It is well known that loop diagrams with internal Higgs exchanges lead to 
a contribution to the dimension-six dipole operators that depends on
products of three Yukawa matrices \cite{Delaunay:2012cz, Beneke:2015lba}. 
This contribution can be sizeable and is an important source of 
flavour violation \cite{Agashe:2006iy}. Therefore it is important to consider 
this {\it Higgs contribution} alongside the previously discussed gauge-contribution.
 
For a bulk Higgs we further need to consider the effect 
of its KK excitations. The mass of the first few Higgs KK states 
is roughly proportional to the inverse width of the corresponding zero-mode 
\cite{Cacciapaglia:2006mz,Davoudiasl:2005uu,Agashe:2014jca}. 
Nonetheless these modes do not necessarily decouple even for a 
strongly localised zero-mode.
This non-decoupling was first shown in \cite{Agashe:2014jca};
the typical impact of the Higgs KK tower is comparable to the effect of 
the zero-mode alone and therefore non-negligible for the 
determination of the dipole operator coefficient.

\begin{figure}
\begin{minipage}{0.45\textwidth}
\includegraphics[width=0.953\textwidth]{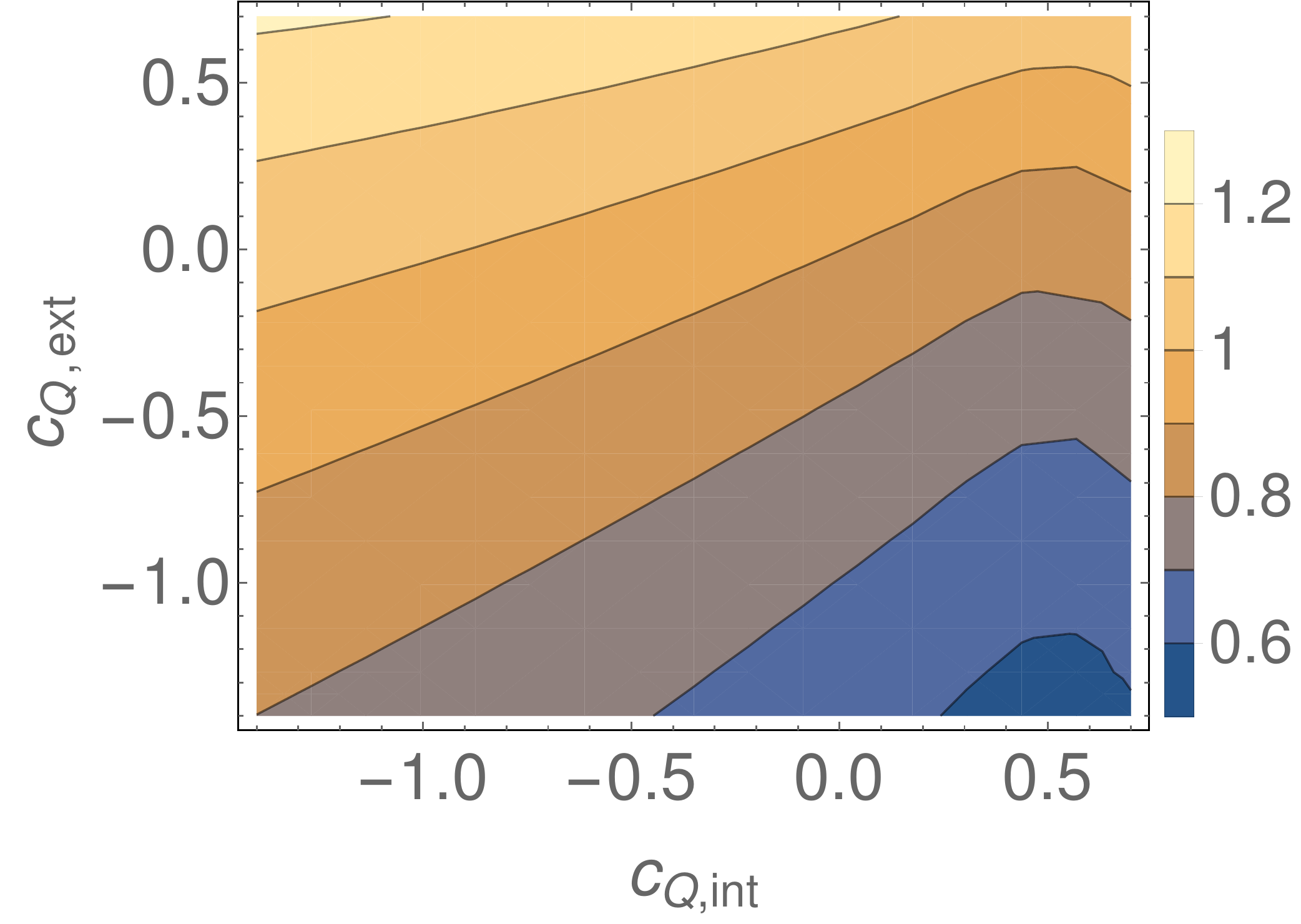} 
\end{minipage}
\hspace{0.5cm}
\begin{minipage}{0.45\textwidth}
\includegraphics[width=0.83\textwidth]{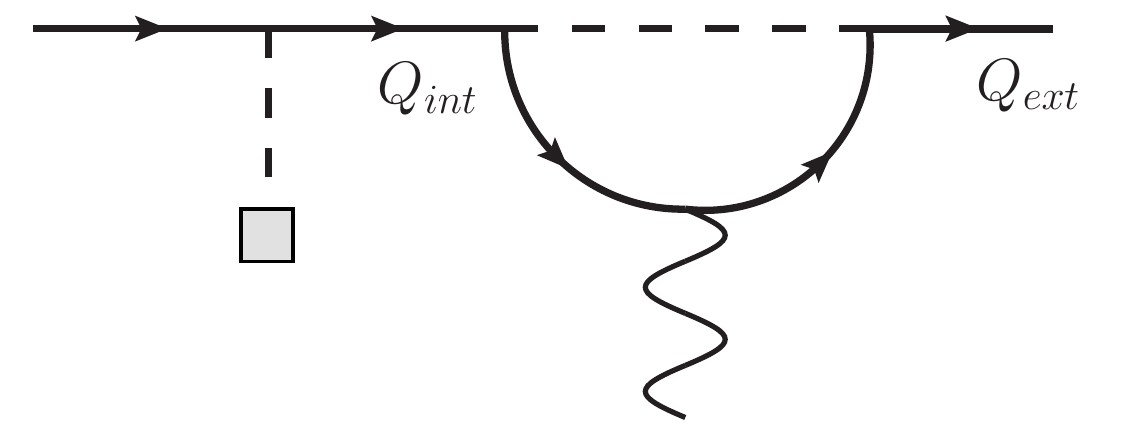}
\end{minipage}
\caption{Illustration the additional dependence of the 5D mass parameter dependence
         of the KK Higgs contribution for the diagram on the right. $c_{Q_{int}}, c_{Q_{ext}}$ are
         the 5D masses of the internal doublet propagator and the external doublet zero-mode.
         For leptons generally only a small region in the upper right corner ($c_Q \sim 0.4-0.7$)
         would be required.
         \label{fig:Massdependence}}.
\end{figure}

Let us first consider the effect of  the zero-mode Higgs only.
We can partially use the results of \cite{Beneke:2015lba} for the leptonic 
dimension-six dipole operator to construct the corresponding
result in the quark sector. Again we only need to replace U(1)
charges and add SU(3) colour factors as appropriate.
For  diagrams where a Higgs is emitted from an external leg and 
not from the loop (see the diagram in figure \ref{fig:Massdependence}
for an example), one further has to distinguish two different
contributions: those where the external quark propagator 
propagates KK modes and so-called off-shell terms that 
arise if the external propagator is a mass-less zero-mode, but the 
$1/p^2$ pole in the propagator is cancelled by powers of $p$ in the numerator,
see \cite{Beneke:2012ie} for a detailed discussion.
The latter terms are basically irrelevant for leptons as
they are effectively suppressed by a SM lepton Yukawa coupling.
They may however play a role in the quark sector due to the 
large top Yukawa coupling and we include these terms in 
$a^{g/\gamma}$.

It is convenient to use the definition $D_\mu=\partial_\mu + i Q_f e A_\mu + i g_s T^A G_\mu$
for the SM covariant derivative with $A_\mu$, $G_\mu$ being photon and gluon field; $e$ is the 
charge of the positron. This definition then coincides with the choice usually employed 
in studies of the $b\to s \gamma$ transition, see e.g.~\cite{Grinstein:1990tj, BurasWeakHam}, and 
makes the expressions in the subsequent sections consistent with the standard literature.

In the minmal RS model we then find
\begin{align}
\label{aphotonmin}
 \left. a_{ij}^\gamma\right|_{Higgs} = & - \frac{ e}{192\pi^2} \frac{T^3}{k^4} 
\, \frac{T^8}{2 k^8} \left( (2Q_e - Q_d - Q_u) F_Q - Q_d F_d + (2 Q_e -Q_u) F_u\right)
\nn \\
& -  \frac{e}{192 \pi^2}  \frac{T^3}{k^4} \,\left(  2 Q_d + Q_u - Q_e  \right) \,  
f^{(0)}_{Q_i}(1/T) 
[Y_d Y^{\dagger}_d Y_d]_{ij} \, g^{(0)}_{d_j}(1/T) \\
\label{agluonmin}
 \left.   a_{ij}^g\right|_{Higgs} = &  \phantom{-}   \frac{ g_s}{192\pi^2} \frac{T^3}{k^4} 
\, \frac{T^8}{2 k^8}  \left( 2 F_Q + F_d+F_u\right)     
\nn \\
& -  \frac{g_s}{192 \pi^2}  \frac{T^3}{k^4} \,3 \,  
f^{(0)}_{Q_i}(1/T) 
[Y_d Y^{\dagger}_d Y_d]_{ij} \, g^{(0)}_{d_j}(1/T)    \,,
\end{align}
where the $F_q$ are abbreviations for
\begin{align}
F_d =&\hphantom{+} f^{(0)}_{Q_i}(1/T) [Y_d]_{ik} F(-c_{{d}_k})
[Y^{\dagger}_d]_{kh}f^{(0)}_{Q_h}(1/T)^2 [Y_d]_{hj} g^{(0)}_{d_j}(1/T) 
\nn \\
 F_Q = &      \hphantom{+} f^{(0)}_{Q_i}(1/T) [Y_{d}]_{ik} g^{(0)}_{d_k}(1/T)^2
[Y^{\dagger}_{d}]_{kh}F(c_{{Q}_h}) [Y_d]_{hj} g^{(0)}_{d_j}(1/T)
 \nn  \\ 
 F_u = &      \hphantom{+}  f^{(0)}_{Q_i}(1/T) [Y_i]_{ik} F(-c_{{u}_k})
[Y^{\dagger}_{u}]_{kh}f^{(0)}_{Q_h}(1/T)^2 [Y_d]_{hj} g^{(0)}_{d_j}(1/T) 
\nn \\
 F_{T_3}= &  \hphantom{+} f^{(0)}_{Q_i}(1/T) [Y_{d}]_{ik} F_{T_3}(c_{{d}_k})
[Y^{\dagger}_{d}]_{kh}f^{(0)}_{Q_h}(1/T)^2 Y_{hj}^{d} g^{(0)}_{d_j}(1/T),
\label{eq:FQuark}
\end{align}
with 
\begin{align}
F(c)\approx & -\frac{k^4}{T^5}\, 
\frac{(3-2c)+(1+2c)\epsilon^{4c-2} - (3-2c)(1+2c)\epsilon^{2c-1} -(1-2 c)^2\epsilon^{1+2c}}
{(1+2 c)(3-2 c)(1-\epsilon^{2c-1})^2 }\,,
\nn \\
F_{T_3}(c) \approx &  -\frac{k^4}{T^5}\, 
\frac{ 1 - \epsilon^{1-2 c}}{1-2 c }\;.
\label{eq:F}
\end{align}
In writing the expression for $F_{T_3}(c)$ we assume that the 
mass parameter $c$ is not too far from $-0.5$, which is realised for 
all parameter points that reproduce the low energy parameters of the SM. 

In the custodially protected model the Higgs contribution to the dipole 
is given by 
\begin{align}
\label{aphotonCS}
  \left. a_{ij}^\gamma\right|_{Higgs} = &-   \frac{ e}{192\pi^2} \frac{T^3}{k^4} 
\, \frac{T^8}{2 k^8} \left((2Q_e - Q_d - Q_u) (F_Q +  F_{T_3})- Q_d F_d + (2 Q_e -Q_u) F_u\right)
\nn \\
& -  \frac{e}{192 \pi^2}  \frac{T^3}{k^4} \,\left( 4 Q_d + 2 Q_u -2 Q_e  \right) \,  
f^{(0)}_{Q_i}(1/T) 
[Y_d Y^{\dagger}_d Y_d]_{ij}\, g^{(0)}_{d_j}(1/T) \\
\label{agluonCS}
 \left.   a_{ij}^g\right|_{Higgs} = &  \phantom{-} \frac{ g_s}{192\pi^2} \frac{T^3}{k^4} 
\, \frac{T^8}{2 k^8} \left(  2 F_Q + 2 F_{T_3} +  F_d + F_u\right)
\nn \\
& -  \frac{g_s}{192 \pi^2}  \frac{T^3}{k^4} \,6 \,  
f^{(0)}_{Q_i}(1/T) 
[Y_d Y^{\dagger}_d Y_d]_{ij}\, g^{(0)}_{d_j}(1/T)    \,.
\end{align}
The terms in \eqref{aphotonmin}, \eqref{agluonmin}, \eqref{aphotonCS} 
and \eqref{agluonCS} with factors of $F_q$, $q=T_3,Q,u,d$, 
correspond to the off-shell contributions.

As already mentioned we also need to take into account the effect of Higgs
KK modes. In \cite{Beneke:2015lba} we absorbed the effect of the KK bosons in global factors 
called $R_i$. These were assumed to be roughly independent of the 5D mass parameters and 
therefore allowed for compact analytic expressions. Nevertheless there is a nontrivial 
dependence of the KK contribution on the 5D mass parameters; in particular for 
diagrams with a Higgs emission from an external line.
In the lepton sector this effect is quite small especially when compared to the sizeable 
numerical uncertainties; we therefore neglected it in \cite{Beneke:2015lba}.
In the quark sector the wide range of 5D masses leads to more
noticeable effects; since we can only determine these numerically 
we do not give an explicit expression. To give an idea of the potential size: 
the left panel in figure \ref{fig:Massdependence} shows the additional effect of the 
mass dependence (without numerical uncertainties) for the diagram 
shown on the right of the same figure.
One can see that the effect is indeed of the order a 
few percent for leptons, but can potentially be of $\mathcal{O}(1)$ for quarks. 
It is therefore not feasible to use a simple analytic approximation as was done
in the lepton sector.

Furthermore, we need to include KK Higgs corrections to the off-shell 
contributions to the Wilson coefficients. 
Again these terms are not necessarily suppressed in the quark sector, as the 
third generation Yukawa couplings are sizeable. However, we can only determine this 
contribution analytically for the Higgs zero-mode and not for the Higgs KK modes; 
it is only accessible numerically, but is quite small, only about $25\%$ of the 
corresponding zero-mode effect.

We therefore treat the effect of whole Higgs KK modes similarly to 
how the gauge-contribution is handled. Here we only remark that 
the total effect of the KK modes is smaller than the effect of the 
Higgs zero-mode but not parametrically so, see also \cite{Agashe:2014jca, Beneke:2015lba}.

\subsection{Beyond QCD}
\label{sec:whatisleftopen}

 We only considered contributions to the Wilson coefficients
that proportional to $\alpha_s$ or, in the case of dipole operators, 
enhanced by 5D Yukawa couplings. Obviously, exchange of hypercharge 
bosons and SU(2) bosons will also generate four-fermion operators,
contribute to both dipoles and give rise to operators of the schematic form
$\Phi^\dagger D_\mu \Phi\; \bar q \gamma^\mu q$. The latter class of operators
will contribute to e.g.~flavour-changing Z couplings.

The U(1) gauge coupling at a scale of $1 \;\rm TeV$ is roughly 
$\alpha_{U(1)}\sim 0.01$. The SU(2)$_L$  coupling is significantly 
larger with  $\alpha_{SU(2)}(\mu =1\;\rm TeV )\sim 0.032$, but still
smaller than $\alpha_s(\mu =1\;\rm TeV )=0.09$. 
The fact that the weak coupling is only about a factor of three
smaller than the strong coupling may warrant including weak effects 
in the matching calculation. Including the effect of the other gauge bosons is not 
a principle problem; their contribution to the four-fermion coefficients
as well as the dipole coefficients can directly be obtained from 
results for leptonic dipoles in the literature, see \cite{Beneke:2015lba}. 

A further effect that would have be taken into account when considering 
weak corrections is the modification of of SM parameters and relations 
that have been utilised in the SM computation. 
In particular the relation of $G_F$ and the $W$ mass, that is frequently used 
when rewriting the SM expressions is affected by higher-dimensional operators 
(see \cite{Alonso:2013hga} for the general case and \cite{Bauer:2009cf}
for the a discussion within the RS model).

It should be noted that KK Higgses do not give rise to 
relevant contributions to the four-fermion operators if the 
Higgs zero-mode is strongly localised towards the IR brane, which we 
always assume. An exchange of a SM Higgs can give a contribution 
to the four-fermion operators. But only in a second
matching step at the intermediate scale $\mu_{int}\sim M_W$ when the 
Higgs degrees of freedom would be removed. 
In this case the flavour-changing Higgs coupling arise from 
dimension-six operators of the form $\bar Q_i \Phi D_j\; \Phi^\dagger \Phi$ 
(see e.g.~\cite{Azatov:2009na}). However, even then the contribution 
will be suppressed by an additional SM b-quark Yukawa coupling. We therefore ignore 
these contributions.

\section{Running to the low scale} 
\label{sec:running} 
The typical energy release in a decay of the type $\bar B\to X_s \gamma$ is of the order of the 
$b$ quark mass and a typical scale choice is thus $\mu_b\ = M_B/2 \approx 2.6 \,{\rm{GeV}}$. From the Standard Model 
calculation of $b \to s \gamma$ in the framework of the weak effective Hamiltonian, 
see \cite{BurasWeakHam} for an overview, it is known that the RGE evolution 
from the weak scale $\mu_W\sim M_W$ down to $\mu_b$ introduces sizeable operator 
mixing \cite{Bertolini:1986tg,Deshpande:1987nr}. 

Our matching calculation was performed the scale $\mu_{KK} \sim T$ and QCD corrections are bound 
to be of importance. We then have two possible strategies: We can either evolve the 
terms in the  dimension-six Lagrangian from the high scale $\mu_{KK}$ to the electroweak scale 
within the unbroken SM, then change to the broken phase and complete the evolution down to the scale $\mu_b$.
The required anomalous dimensions for the first step can be found e.g.~in \cite{Alonso:2013hga,Jenkins:2013zja,Jenkins:2013wua}.
Alternatively, we can work with the ``broken" operator basis already at the high scale and 
perform the evolution down to the low scale in one step (taking into account the top-mass
threshold). The first approach is more in the spirit of a matching onto a set of dimension-six operators. 
The second option has simpler ``logistics" as we only need consider a single RGE.
Both strategies are valid and ultimately must be equivalent in a situation where
no additional dynamics between $\mu_{KK}$ and $\mu_W$ need to be taken into account.

However, for the specific process at hand the second option has the 
additional advantage that the structure of the required evolution equation
has been studied in some detail in \cite{Buras:2011zb}. While \cite{Buras:2011zb} ultimately focusses on
scenarios with e.g.~a flavour-changing  $Z^\prime$, their operator basis 
contains the full set of normal and colour-flipped four-quark operators.
We therefore choose to follow this approach. 

Let us for clarity introduce the  effective Hamiltonian at the 
high scale $\mu_{KK}$, that is used in \cite{Buras:2011zb}
\begin{align}
\label{EWHam}
 \mathcal{H}^{(b\to s)}=-\frac{4 G_F}{\sqrt{2}}& V^\star_{ts} V_{tb}
        \bigg[  \Delta C_{7\gamma}(\mu_{KK})Q_{7\gamma} +\Delta C_{8g}(\mu_{KK})Q_{8g} 
              + \Delta C^\prime_{7\gamma}(\mu_{KK})Q^\prime_{7\gamma} +\Delta C^\prime_{8g}(\mu_{KK})Q^\prime_{8g} 
  \nn \\  
             & + \sum_{A,B=L,R}\sum_{q=u,c,t,d,s,b} \Delta C^{q}_{1}[A,B](\mu_{KK})\,Q^{q}_{1}[A,B] 
             +\Delta C^{q}_{2}[A,B](\mu_{KK})\;Q^{q}_{2}[A,B] 
  \nn \\  
              &+ \sum_{A,B=L,R} \Delta \widehat{C}^{d}_{1}[A,B](\mu_{KK})\,\widehat{Q}^{d}_{1}[A,B] 
              +\Delta \widehat{C}^{d}_{2}[A,B](\mu_{KK})\;\widehat{Q}^{d}_{2}[A,B] 
 \bigg]
\end{align}
where the operators are given by 
\begin{align}
\label{Operators_New}
Q_{7\gamma}&=\frac{e\,m_b}{16 \pi^2} \bar s_\alpha \sigma^{\mu\nu} P_R b_\alpha F_{\mu\nu}&
Q_{8g}&=\frac{g_s\,m_b}{16 \pi^2} \bar s_\alpha \sigma^{\mu\nu} P_R T^A_{\alpha\beta} b_\beta G^A_{\mu\nu}
\nn\\
Q^\prime_{7\gamma}&=\frac{e\,m_b}{16 \pi^2} \bar s_\alpha \sigma^{\mu\nu} P_L b_\alpha F_{\mu\nu}&
Q^\prime_{8g}&=\frac{g_s\,m_b}{16 \pi^2} \bar s_\alpha \sigma^{\mu\nu} P_L T^A_{\alpha\beta} b_\beta G^A_{\mu\nu}
\nn\\
Q^{q}_{1}[A,B]&=(\bar s_\alpha \gamma^\mu P_A b_\beta)\;(\bar q_\beta \gamma_\mu P_B q_\alpha) &
Q^{q}_{2}[A,B]&=(\bar s_\alpha \gamma^\mu P_A b_\alpha)\;(\bar q_\alpha \gamma_\mu P_B q_\alpha) 
\nn\\
\widehat{Q}^{d}_{1}[A,B]&=(\bar s_\alpha \gamma^\mu P_A d_\beta)\;(\bar d_\beta \gamma_\mu P_B b_\alpha) &
\widehat{Q}^{d}_{2}[A,B]&=(\bar s_\alpha \gamma^\mu P_A d_\alpha)\;(\bar d_\alpha \gamma_\mu P_B b_\alpha) 
\end{align}
with $P_{L/R}=\frac12 (1\mp\gamma_5)$ as usual and $\alpha,\,\beta$ are colour indices.
Note that while the usual current-current and penguin operators
\begin{align}
\label{Operators_SM}
Q_1&= (\bar s_\alpha \gamma^\mu P_L c_\beta)\;(\bar c_\beta \gamma_\mu P_L b_\alpha)&
Q_2&= (\bar s_\alpha \gamma^\mu P_L c_\alpha)\;(\bar c_\beta \gamma_\mu P_L b_\beta)
\nn \\
Q_3&= (\bar s_\alpha \gamma^\mu P_L c_\alpha)\!\!\!\sum_{q=u,c,d,s,b}(\bar q_\beta \gamma_\mu P_L q_\beta)&
Q_4&= (\bar s_\alpha \gamma^\mu P_L c_\beta)\!\!\!\sum_{q=u,c,d,s,b}(\bar q_\beta \gamma_\mu P_L q_\alpha)
\nn \\
Q_5&= (\bar s_\alpha \gamma^\mu P_L c_\alpha)\!\!\!\sum_{q=u,c,d,s,b}(\bar q_\beta \gamma_\mu P_R q_\beta)&
Q_6&= (\bar s_\alpha \gamma^\mu P_L c_\beta)\!\!\!\sum_{q=u,c,d,s,b}(\bar q_\beta \gamma_\mu P_R q_\alpha)
\end{align}
are not included in \eqref{EWHam}, they do enter the renormalisation group equations. 
This operator basis is obviously non-minimal as e.g.~$Q_1$ and $Q^c_2[L,L]$ are related
via Fierz identities. As we only consider the LO corrections due to new physics, this does not 
invalidate the RG analysis \cite{BurasWeakHam}.  

In total we have to consider $70$ operators. Fortunately, there are only a few 
independent entries in the leading order (LO) anomalous dimension 
matrix. Most of which can be taken from \cite{Ciuchini:1993ks, Ciuchini:1993fk} once the different 
operator normalisation has been taken into account\footnote{In \cite{BurasWeakHam} 
the corresponding operators $Q_{1-8}$ are only rescaled by a factor of $1/4$ compared to their 
definition in \eqref{Operators_New},\eqref{Operators_SM}. The anomalous dimensions remain
therefore the same.}. 
The remaining entries can be taken directly from \cite{Buras:2011zb} where the
use of effective, scheme-independent coefficients $C_{7\gamma}^{eff}$, $C_{8g}^{eff}$
is implied. In the following we tacitly assume that $C_{7/8}$ refers to the effective quantity
and forgo to display the superscript.
We will not give the anomalous dimensions explicitly and refer to the original literature for 
details. 

With the anomalous dimensions at hand, the renormalisation group evolution equation (RGE)
\begin{align}
\label{eq:RGE}
 \mu \frac{d}{d\mu} \vec{C}_i(\mu)=\frac{\alpha_s(\mu)}{4\pi} [\gamma^T]_{ij} \vec{C}_j(\mu)
\end{align}
can be solved in the standard way, provided the initial conditions 
at the high scale $\mu_{KK}$ are known. As the anomalous dimension matrix 
$\gamma$ is sparse, a basis where the evolution is diagonal can be determined
very efficiently. For the strong coupling constant we use $\alpha_s(M_Z)=0.1185$
with decoupling of the top quark at $m_t=170\,{\rm GeV}$.

Once the evolution down to $\mu_b$ has been performed the result for the branching fraction 
of $\bar B \to X_s \gamma$ can be obtained using the formula \cite{Buras:2011zb, Niehoff:2015iaa}
\begin{align}
\frac{\left. \text{Br}\left( B \to X_s \gamma\right)\right|_{E_\gamma>1.6\,{\rm GeV}}}
{\left. \text{Br}\left( B \to X_s \gamma\right)\right|^{SM}_{E_\gamma>1.6\,{\rm GeV}}}= 
\frac{1}{\left|C_{7\gamma}(\mu_b)^{SM}\right|^2 +N}
      \left( \left|C_{7\gamma}(\mu_b)\right|^2   + \left|C^\prime_{7\gamma}(\mu_b)\right|^2  +N\right)\;.
\end{align}
Here we use a minimum photon energy of $E_\gamma^{min}=1.6\,{\rm GeV}$; the same as was used for 
the HFAG world average. Here the $N$ is a non-perturbative 
correction \cite{Falk:1993dh,Gambino:2001ew,Bauer:1997fe,Neubert:2004dd} 
and we use $N(E_\gamma=1.6\;{\rm GeV})=3.6\times 10^{-3}$.

Since we work in leading order in the new physics contribution, BSM effects only induce
a shift in the Wilson coefficients
\begin{align}
 C^{(\prime)}_{7\gamma}(\mu_b)\to \left[C^{(\prime)}_{7\gamma}(\mu_b)\right]_{SM} +\Delta C^{(\prime)}_{7\gamma}(\mu_b)\;.
\end{align}
The SM value of the dipole coefficients
\begin{align}
 C_{7\gamma}(\mu_b)= -0.368
\end{align}
can be taken from \cite{Misiak:2015xwa}. The primed coefficient $C_{7\gamma}$ is 
tiny as it is suppressed by $m_s/m_b$ and can be neglected.

For completeness we also give the formulae for the
related process $b\to s g$. It can be treated completely
analogously; here the shifts in the coefficients $C^{(\prime)}_{8g}(\mu_b)$ are required.
The NLL SM prediction was determined in \cite{Greub:2000sy}.
The partial width for the process $b\to s g$ is given by 
\begin{align}
 \Gamma\left( \bar B  \to s g \right)= \frac{\alpha_s(m_b) m_b^5}{24 \pi^2} 
 \left|G_F V_{ts}^\star V_{tb} \right|^2 |D|^2 + \Gamma_{fin}^{\text{brems}}\;.
\end{align}
The explicit expressions for $\Gamma_{fin}^{\text{brems}}$ and $D$ can be found in \cite{Greub:2000sy}.
The branching fraction is then obtained as 
\begin{align}
 {\text{Br}}(b\to s g) \approx \frac{\Gamma(b\to s g)}{\Gamma(b\to c e \bar{\nu}_e)} \mathcal{B}^{exp}_{sl}\;.
\end{align}
With $\mathcal{B}^{exp}_{sl}\approx 0.105\pm 0.005$ being the experimental semi-leptonic branching fraction 
of the B-meson. In the SM one finds \cite{Greub:2000sy}
\begin{align}
{\text{Br}}^{SM}(b\to s g)= (5.0 \pm 1.0)\times 10^{-3}\;.
\end{align}

The last missing piece for our analysis are the initial conditions
for the RGE. That is, we need the Wilson coefficients $\Delta C$ in 
\eqref{EWHam}.

\subsection*{Initial conditions}

The Wilson coefficient in $\mathcal{H}^{b\to s}$ at the high scale $\mu_{KK}$
can be obtained from the Wilson coefficients of the dimension-six operators
in \eqref{eq:LagraniganHigh}.
We need to rotate into the low-energy mass basis and replace
the SM Higgs field (if present) by its vacuum expectation value $v/\sqrt{2}$:
\begin{align} 
 \label{SubstitutionForBrokenPhase} 
   \Phi \to \begin{pmatrix} 
             \phi^+\\ 
             \frac{1}{\sqrt{2}}(v+h+iG) 
            \end{pmatrix} && 
    Q_i \to   P_L\begin{pmatrix} 
             U^u_{ij}u_j\\ 
             U^d_{ij}d_j 
            \end{pmatrix} && 
    U_i\to V^u_{ij}P_R u_j &&   
    D_i\to V^d_{ij}P_R d_j \;.
\end{align} 
We only take into account terms that contribute to the Wilson 
coefficients in \eqref{EWHam} and drop all others.

As an example, let us consider the term $b^{DU}_{ij}  \bar D_i  \gamma^\mu T^a D_i  \bar U_j \gamma^\mu T^a U_j$ in the 
dimension-six Lagrangian. 
Using the substitution rules \eqref{SubstitutionForBrokenPhase} we find
\begin{align}
\label{bDUconversion}
  b^{DU}_{ij}  \bar D_i& \gamma^\mu T^A D_i  \bar U_j \gamma_\mu T^A U_j \longrightarrow
  \beta^{DU}_{sbq_uq_u}\, \bar s \gamma^\mu T^A P_R b\, \bar q_u \gamma_\mu T^A P_R q_u\, = \nn \\
 &=  -\frac {1}{2 N_c} \beta^{DU}_{sbq_uq_u}\, \bar s \gamma^\mu  P_R b\, \bar q_u \gamma_\mu  P_R q_u
+\frac12  \beta^{DU}_{sbq_uq_u}\, \bar s_\alpha \gamma^\mu  P_R b_\beta\, (\bar q_u)_\beta \gamma_\mu  P_R (q_u)_\alpha\nn \\
 & =-\frac {1}{2 N_c} \beta^{DU}_{sbq_uq_u}\, O_2^{q_u}[R,R]
  +\frac12  \beta^{DU}_{sbq_uq_u}\, O_1^{q_u}[R,R]
\end{align}
where a simple single sum over $q_u=u,c,t$ is implied. Here we defined
$\beta^{DU}_{sbq_u q_u} =[V^d]^\dagger_{si}[V^u]^\dagger_{q_uj} b^{DU}_{ij} V^d_{ib} V^u_{jq_u}$.
In general  we will use the abbreviation
\begin{align}
 \beta^{FF^\prime}_{ABCD}=[R^F]^\dagger_{Ai}[R^{F^\prime}]^\dagger_{Cj} b^{F F^\prime}_{ij} R^F_{iB} R^{F^\prime}_{jD}
\end{align}
with the appropriate flavour rotation matrices $R^{F^{(\prime)}}$.

Comparing \eqref{bDUconversion} with \eqref{EWHam}, we obtain 
\begin{align}
\frac{4 G_F V^\star_{ts} V_{tb}}{\sqrt{2}} \Delta C^{q_u}_{1}[R,R](\mu_{KK})=\frac{1}{2 T^2} \beta^{DU}_{sbq_uq_u}, &&
\frac{4 G_F V^\star_{ts} V_{tb}}{\sqrt{2}} \Delta C^{q_u}_{2}[R,R](\mu_{KK})=-\frac{1}{2N_c T^2} \beta^{DU}_{sbq_uq_u}\,.
\end{align}
The remaining four-quark operators can be related to operators in the weak Hamiltonian 
in the same fashion. For clarity, we have relayed the expressions for the Wilson coefficient of \eqref{EWHam}
to an Appendix. 

Similarly, we can obtain the effective dipole operator coefficients. 
Introducing the abbreviation $\alpha^{\gamma/g}=U_d^\dagger a^{\gamma/g}V_d$ we find
\begin{align}
 \frac{4 G_F V^\star_{ts} V_{tb}}{\sqrt{2}}  C_{7\gamma}(\mu_{KK})&=  \frac{16 \pi^2}{e\,m_b\,T^2} \phantom{[}\alpha^\gamma_{sb}\phantom{]^\dagger}\frac{v}{\sqrt{2}} 
                         + \sum_{q=d,s,b} \frac{Q_q \, m_q\, C_F}{m_b\,T^2} \beta^{QD}_{qbsq}        \nn \\
 \frac{4 G_F V^\star_{ts} V_{tb}}{\sqrt{2}} C^{\prime}_{7\gamma}(\mu_{KK})&=  \frac{16 \pi^2}{e\,m_b\,T^2} [\alpha^\gamma]^\dagger_{sb}\frac{v}{\sqrt{2}} 
                         + \sum_{q=d,s,b} \frac{Q_q  \, m_q\, C_F}{m_b\,T^2} \beta^{QD}_{sqqb}            \nn \\
\frac{4 G_F V^\star_{ts} V_{tb}}{\sqrt{2}}  C_{8g}(\mu_{KK})&=   \frac{16 \pi^2}{g_s\, m_b \,T^2} \phantom{[}\alpha^g_{sb}\phantom{]^\dagger}\frac{v}{\sqrt{2}} 
                         - \sum_{q=d,s,b} \frac{ m_q}{2N_c\, m_b\, T^2} \beta^{QD}_{qbsq}                   \nn \\
 \frac{4 G_F V^\star_{ts} V_{tb}}{\sqrt{2}} C^{\prime}_{8g}(\mu_{KK})&= \frac{16 \pi^2}{g_s\, m_b \,T^2} [\alpha^g]^\dagger_{sb}\frac{v}{\sqrt{2}} 
                         - \sum_{q=d,s,b} \frac{m_q}{2N_c \, m_b\,T^2} \beta^{QD}_{sqqb}              \;.
                         \label{eq:Cdipoles}
\end{align} 
All quantities on the right-hand side of \eqref{eq:Cdipoles} are implied to be evaluated at the scale $\mu_{KK}$.
The terms containing a $\beta$-coefficient arise from the one-loop diagrams with an insertion of a four-fermion operator.
They ensure that the (effective) coefficient $\Delta C_{7\gamma}$ is scheme independent, see Sec.~\ref{sec:dipole}.

\section{Phenomenology} 
\label{sec:pheno} 

\begin{figure}
 \includegraphics[width=0.45\textwidth]{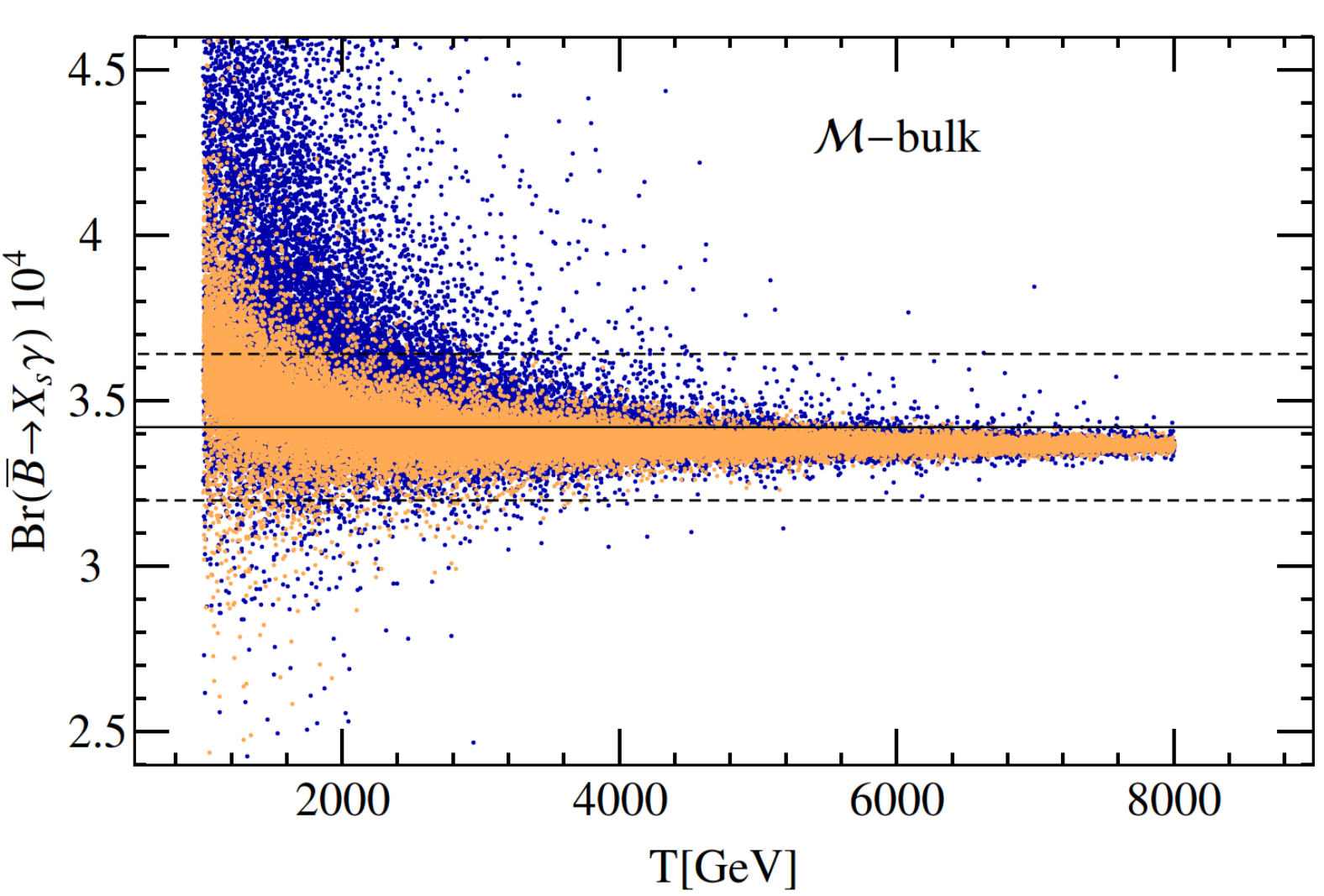}\hspace{1.2cm}
 \includegraphics[width=0.45\textwidth]{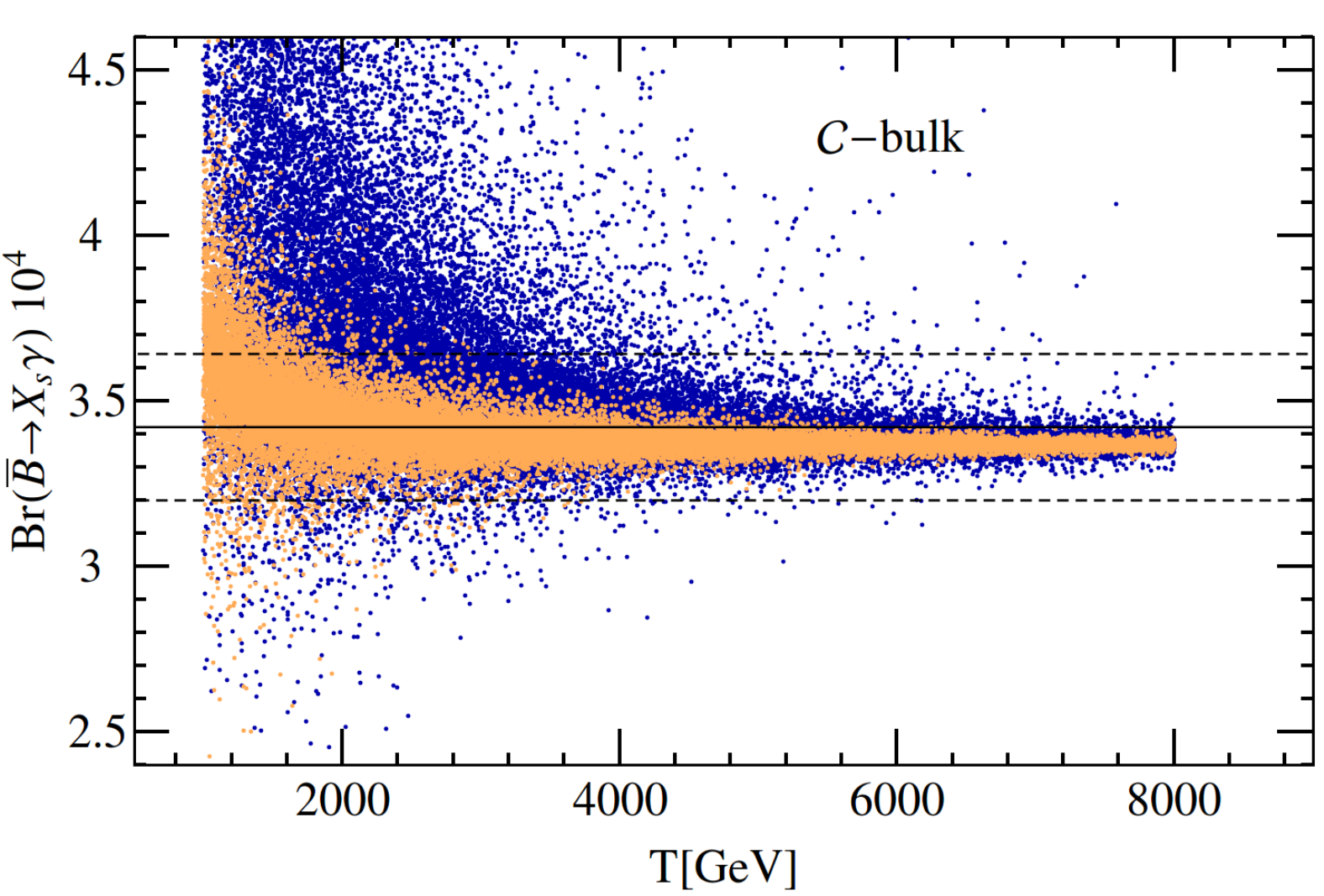}
 \caption{  ${\rm{ Br}}(\overline{B} \to X_s \gamma)$ as a function of the KK scale $T$. The blue (dark grey) points correspond to 
 the data set with large Yukawas, $Y_{max}=3$. The orange (light grey) points correspond to $Y_{max}=1/2$. The horizontal lines indicate the 
 experimental value of and uncertainty on the branching fraction.  The left panel shows the result for the minimal RS model, the right panel for the
 custodially protected model.
  \label{fig:bsgVsT} }
\end{figure}

To see the potential effect of the additional contribution
to $C_{7\gamma}^{(\prime)}$ on the $B\to X_s \gamma$ decay we need to scan 
over the parameter space of the RS model. We will, as mentioned before, consider
a minimal and a custodially protected RS model with an IR-localised bulk Higgs.
The model parameters  include the 5D masses of the fermions as well as
the two Yukawa couplings $Y_d$ and $Y_u$.
These parameters are not independent as we need to impose the condition that
low-energy parameters of the SM are reproduced within uncertainties.
We take into account the SM quark masses (at the scale $T$) and the CKM 
angles and phase; here we make use of the analytic approximations
of \cite{Casagrande:2008hr}. A further restriction is imposed by hand on the dimensionless
Yukawa matrices as we require them to be anarchic. That is, the matrix elements 
all have roughly a common magnitude of $\mathcal{O}(1)$ and arbitrary phase.
Similar to the analysis in \cite{Beneke:2015lba} we consider two samples of 
Yukawas: one with a maximum entry size of $Y_{max}=3$ (representing the case of
large Yukawa couplings) and one with an upper bound of $Y_{max}=1/2$ 
(representing the case of small Yukawa couplings).

The main result of our scan through the RS parameter space is shown in 
figure \ref{fig:bsgVsT}. It shows the branching fraction $\bar B \to X_s \gamma$
as a function of the KK scale\footnote{Note that the 
mass of the first KK excitation of the gluon is roughly given by $2.5 \times T$ 
\cite{Pomarol:1999ad}} $T$ for the minimal RS model (left panel) and the custodially protected model 
(right panel). The blue (dark grey) points correspond to 
$Y_{max}=3$, the orange (light grey) points to $Y_{max}=1/2$. The current 
experimental central value, see equation \eqref{eq:bsgammaExp}, is represented by the solid 
horizontal line; the dashed lines indicate the uncertainty. 

We find that the branching fraction is, especially for small Yukawas, predominantly larger 
than in the SM. This is due to a sizeable contribution 
from $C_{7\gamma}^{\prime}$, that 
lacks an unsuppressed interference term with the SM contribution---its contribution 
to the branching fraction is always positive. In addition to that the 
contribution to $C_{7\gamma}^{\prime}$ is generally larger than 
the contribution to the unprimed dipole coefficient. The reason for this, as was observed already in 
\cite{Blanke:2012tv}, is that the 5D profile of the doublet $Q_3$ (that
very roughly corresponds to the $b_L$ after EWSB) is typically larger 
than the profiles of the down-type singlets $D$ near the IR brane; 
consequently the operator $Q_{7,\gamma}^\prime\propto  
(\overline{s_R})_\alpha \sigma^{\mu\nu} (b_L)_\alpha F_{\mu\nu} $ receives a 
larger BSM contribution.

Only for the $Y_{max}=3$ sample one can observe data points with a significantly
reduced branching fraction compared to the SM. This is due to a destructive interference of 
$C_{7\gamma}^{SM}$ and $\Delta C_{7\gamma}$ that can counteract the contribution due to $C_{7\gamma}^\prime$
if the Higgs contribution to $C_{7,\gamma}(\mu_{KK})$ is large. This effect is more 
pronounced in the custodially protected model where the additional fermion states 
enhance the dipole coefficient, cp.~\eqref{aphotonmin} and \eqref{aphotonCS}.
For small Yukawas the phenomenology of minimal and custodially protected model is
quite similar. This is to some extent a consequence of working only with QCD- and Higgs-mediated 
contributions to the Wilson coefficient; QCD is treated the same in both models 
while the electroweak sector is extended and features additional bosonic modes.
In the $Y_{max}=1/2$ scenario the main distinction between the two models---the Higgs contribution---
is suppressed.

\begin{figure}
 \includegraphics[width=1\textwidth]{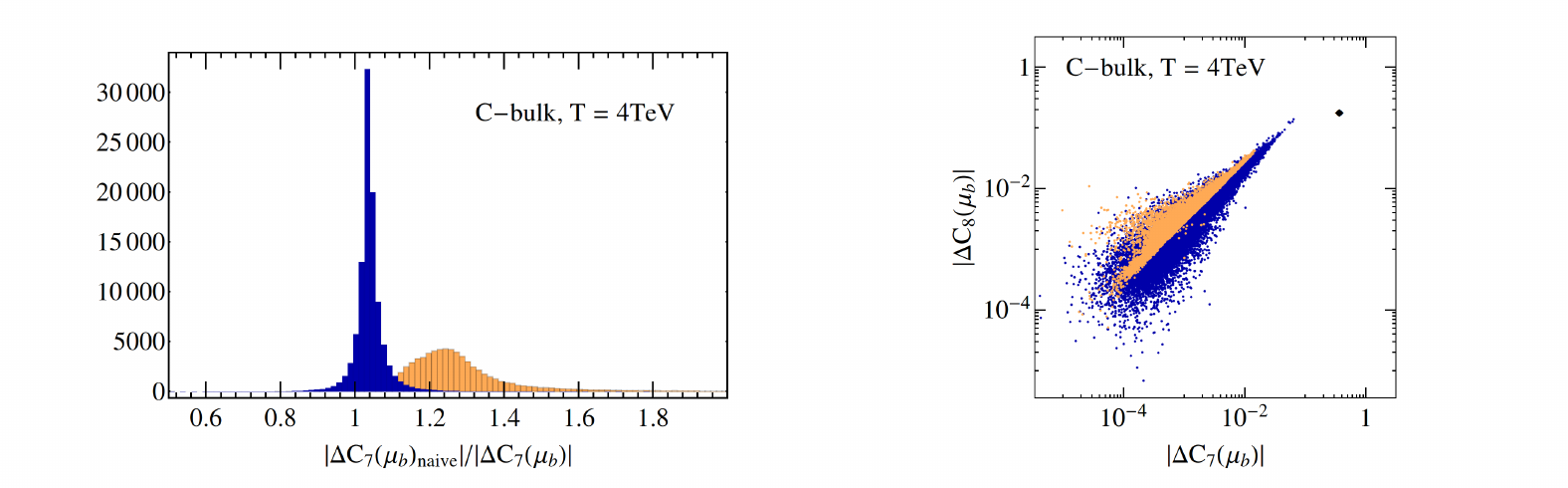} 
 
 \caption{ {\it left panel:} Effect of operator mixing on $\Delta C_{7\gamma}$. The 
 histogram shows the distribution of $|\Delta C_{7\gamma}(\mu_b)|$ without operator mixing
 relative to the full   $|\Delta C_{7\gamma}(\mu_b)|$ with mixing (see text for details).
  The blue (dark grey) and orange (light grey) histogram corresponds to  $Y_{max}=3$  and $Y_{max}=1/2$.  
 {\it right panel:} Correlation of $\Delta C_{8g}$ and $\Delta C_{7\gamma}$ in the custodially protected
                    RS model for $T=4\,{\rm TeV}$. The triangle represents the SM values of $C_{7\gamma}$
                    and $C_{8g}$; the dashed diagonal line indicates $|\Delta C_{8g}|=|\Delta C_{7\gamma}|$. 
                    Same colour coding as in the left panel. 
  \label{fig:Mixing} }
\end{figure}

The smallness of the Higgs contribution for $Y_{max}=1/2$ and the consequently 
smaller $\Delta C_{7\gamma}(\mu_{KK})$ also make the inclusion of operator mixing 
mandatory. To see this we consider two quantities: the full $\Delta C_{7\gamma}(\mu_{b})$
as obtained from the RGE \eqref{eq:RGE} and $\Delta C_{7\gamma}(\mu_{b})|_{naive}$
which is also obtained via \eqref{eq:RGE} but we set the Wilson coefficients
of all four-fermion operators at the high scale $\mu_{KK}$ to zero.
We then consider the ratio $\Delta C_{7\gamma}(\mu_{b})|_{naive}/\Delta C_{7\gamma}(\mu_{b})$.
The deviation of the ratio from one indicates the relative importance of the 
four-fermion operators for the $b\to s\gamma$ transition. 
Histograms of $\Delta C_{7\gamma}(\mu_{b})|_{naive}/\Delta C_{7\gamma}(\mu_{b})$ are 
shown in the left panel of figure \ref{fig:Mixing}. For simplicity we only show the 
plot in the minimal model for $T=4\,{\rm TeV}$. For large Yukawas, $Y_{max}=3$ in blue (dark grey),
neglecting the contribution of from four-fermion operators leads on average 
to an increase of $\Delta C_{7\gamma}(\mu_{b})$ by $5\%$. For a few Yukawa data sets 
the shift can be of the order of $\pm 15 \%$. In the case of small 5D Yukawa coupling (shown in orange)
ignoring the four-fermion operator mixing basically always increases $\Delta C_7$. This can lead 
to an overestimate of the BSM contribution to the $\bar B\to X_s \gamma$ branching 
fraction by up to $40\%$.
Hence including the mixing is relevant and should not be neglected. This is of 
course quite general as FCNCs mediated by new, massive gauge bosons usually 
create simultaneous contributions to $\Delta C_{7\gamma}$ and to the $\Delta C_{1,2}^{q}[A,B]$
as is indicated by the need to include the four-fermion operators to obtain a 
scheme-independent result.

For completeness we also show the correlation of $\Delta C_{7\gamma}(\mu_{b})$ and 
$\Delta C_{8\gamma}(\mu_{b})$ in the right panel of figure \ref{fig:Mixing}.
We see that on average the BSM contribution to $C_{7\gamma}$ is 
smaller than the contribution to $C_{8g}$ as was also noted in \cite{Blanke:2008zb}. 
This is more noticeable for the small Yukawa sample shown in orange (light grey).
The two  Wilson coefficients are then clearly correlated and one observes a "lower bound" 
on  $\Delta C_{7\gamma}$ for a given value of $\Delta C_{8\gamma}$.
However, with $Y_{max}=3$, it is straightforward to find parameter points 
where $\Delta C_{7\gamma}$ is much larger than the BSM contribution to $C_{8g}$.
The reason for this is the following:
The zero-mode Higgs contribution to $a^g$ and $a^\gamma$ are almost proportional to each other, see 
equations \eqref{aphotonmin}--\eqref{aphotonCS}. However, the sizeable KK Higgs 
contribution has a more complicated structure; it contributes in a different way to
$a^g$ and to $a^\gamma$. This blurs the correlation.

Finally, comparing with the experimental value for $\bar B \to X_s \gamma$ we find that 
for $Y_{max}=1/2$ the RS model parameter space is generally compatible with experimental 
data for $T>2\;{\rm TeV}$. Since electroweak precision observables already put stricter bounds on
the KK scale \cite{Casagrande:2008hr,Casagrande:2010si}, $\bar B \to X_s \gamma$ does not 
give any new constraints on the KK scale. Nonetheless, sizeable corrections
of about $5-10\%$ are still possible.    
For large Yukawas the situation is much more intriguing, especially in the custodially
protected model. As the large effects come almost 
exclusively from the Higgs exchange contribution to the dipole coefficients $a^{g/\gamma}$
they are strongly dependent on the specific form of the anarchic Yukawa matrices.
It is difficult to deduce any hard bounds on the RS parameter space. However, the total
BSM correction to the branching fraction can be quite substantial. 
Even for $T\sim 5 \;{\rm TeV}$ it is easy to find parameter points outside
the current experimental limits. Consequently, the new Belle II searches would
have the potential to discover the impact of KK states on $\bar B\to X_s\gamma$
with masses well above $10\;{\rm TeV}$. The search at the B-factory is therefore complementary to other 
 powerful indirect search avenues like Higgs production and decay
 or dipole moments --- experiments at vastly different energy scales.

\section{Conclusion} 
\label{sec:conclusion} 

We have studied the flavour violating radiative transition $b\to s \gamma$
in RS models with an IR localised bulk Higgs. For simplicity, our analysis 
is restricted to QCD- and Higgs-mediated BSM effects.
We followed the strategy of \cite{Beneke:2012ie} and matched the five-dimensional 
RS model onto the SM effective theory including dimension-six operators.
Here we could make use of our recent results \cite{Beneke:2015lba} for 
lepton-flavour violation in the RS model. In particular 
the complicated 5D loop integrals that determine the dimension-six quark dipole 
coefficients could be recovered from the electromagnetic dipole coefficient for 
leptons. This way we can include the effect of 5D loops with internal gauge,
Higgs as well as KK Higgs bosons.

After the transition to the broken electroweak phase, we used the results 
of \cite{Buras:2011zb} to include the effect of operator mixing due to 
RGE evolution from the KK scale $T$ to $\mu_b$ to LL accuracy.
This is necessary as already in the SM the QCD corrections are sizeable and 
the dipole operator coefficient alone is not regularisation scheme 
independent. We find that for small Yukawa couplings, i.e., for
small Higgs contributions to the dimension-six dipoles, the mixing of 
additional four-fermion Wilson coefficients into $C_{7,8}$
can be sizeable and should not be neglected. We expect this to 
be true in any BSM model where dipole and four-fermion operators are generated
via exchange of the same intermediate states.

While our results for the Wilson coefficients are general, we 
assumed anarchic Yukawa couplings to study the phenomenology of 
the decay $\bar B\to X_s \gamma$ in both the minimal and the 
custodially protected RS model. We find that the additional contributions to 
the branching fraction can be sizeable for large Yukawas
and moderate KK scales $T$.

The strong sensitivity of the RS contribution to the specific form of the Yukawa matrices
makes it challenging to directly constrain the parameter space of the model.
Nonetheless, the decay is a useful tool that complements other powerful probes for the 
KK scale in the quark sector, like Higgs production/decay
\cite{Malm:2013jia,Malm:2014gha,Archer:2014jca}. 
More importantly, for large 5D quark Yukawas there can be observable deviations
of ${\rm Br}(\bar B\to X_s \gamma)$ from its SM value even for masses of the first KK excitation 
of around $10\;{\rm TeV}$.
For small 5D Yukawas couplings ($Y_{max}\sim 0.5$) the impact of the RS model is mild; 
for KK scales that are not in conflict with electroweak 
precision measurements the $\bar B \to X_s\gamma$ branching fraction generally agrees 
with the current world average within uncertainties. In this case the aforementioned alternative search channels are more promising.

\paragraph{Note added:} 
While this work was in its final stage, \cite{Malm:2015oda} was published.
\cite{Malm:2015oda} presents a detailed analysis of the $b\to s \gamma$
transition in the minimal RS model with an exactly brane-localised Higgs. It is to our knowledge also the first 
computation of the RS contribution to dipole operators that does not rely 
on an expansion in the ratio of electroweak and KK scale. In addition to QCD and Higgs 
effects also electroweak effects are taken into account, but the model does, by construction, not involve Kaluza Klein
Higgs contributions.   \cite{Malm:2015oda} includes QCD operator mixing, but neglects the effect of the four-fermion operators. 
Since we consider the case of a localised bulk Higgs with KK modes, it is most 
useful to compare with the case of small Yukawa couplings; in this case the quite different Higgs sector
does not play an all too dominant role. We then find RS corrections to ${\rm Br}(\bar B \to X_s\gamma)$
that are of similar but slightly smaller in size to those found \cite{Malm:2015oda}.
This seems not unexpected as we neglect electroweak corrections to the dipole, but do include mixing with dimension-six 
fermion operators, which tends to give rise to a slightly smaller $\Delta C_{7\gamma}$ coefficient.

\paragraph{Acknowledgements:} 
We are grateful to M.~Beneke for suggesting this project and for many useful discussions.
The work of P.M.~is supported in part by the Gottfried Wilhelm Leibniz programme 
of the Deutsche Forschungsgemeinschaft (DFG). The work of J.R.~is supported by STFC UK. 
We thank the Munich Institute for Astro- and Particle Physics (MIAPP) of 
the DFG cluster of excellence ``Origin and Structure of the Universe'' for 
hospitality during part of the work.
The Feynman diagrams were drawn with the help of Axodraw \cite{Vermaseren:1994je} and 
JaxoDraw \cite{Binosi:2003yf}. 

\appendix

\section{Wilson coefficients of the extended electroweak Hamiltonian at the scale $\mu_{KK}$}
\label{AppendixWCs}
In the following we collect the coefficients of the various four-fermion operators 
in \eqref{EWHam}. To this end we first map each operator in the dimension-six
Lagrangian unto operators in the broken electroweak theory and extract the 
Wilson coefficients by comparing with \eqref{EWHam}. For brevity, let us first introduce the abbreviation $\mathcal{V}= \frac{4 G_F V^\star_{ts} V_{tb}}{\sqrt{2}} $. 

\begin{align}
  b^{QU}_{ij}  \bar Q_i& \gamma^\mu T^A Q_i  \bar U_j \gamma_\mu T^A U_j \longrightarrow
  \beta^{QU}_{sbq_uq_u}\, \bar s \gamma^\mu T^A P_R b\, \bar q_u \gamma_\mu T^A P_R q_u\, = \nn \\
 &=  -\frac {1}{2 N_c} \beta_{sbq_uq_u}\, \bar s \gamma^\mu  P_L b\, \bar q_u \gamma^\mu  P_R q_u
+\frac12  \beta_{sbuu}\, \bar s_\alpha \gamma^\mu  P_L b_\beta\,  (\bar q_u)_\beta \gamma^\mu  P_R (q_u)_\alpha\nn \\
 & =-\frac {1}{2 N_c} \beta_{sbq_uq_u}\, O_2^{q_u}[L,R]
  +\frac12  \beta_{sbq_uq_u}\, O_1^{q_u}[L,R]
\end{align}
gives 
\begin{align}
\mathcal{V} \Delta C^{q_u}_{1}[L,R](\mu_{KK})=\frac{1}{2 T^2} \beta^{QU}_{sbq_uq_u} &&
\mathcal{V} \Delta C^{q_u}_{2}[L,R](\mu_{KK})=-\frac{1}{2N_c T^2} \beta^{QU}_{sbq_uq_u}\,.
\end{align}

\begin{align}
  b^{DD}_{ij}  \bar D_i& \gamma^\mu T^A D_i  \bar D_j \gamma_\mu T^A D_j \longrightarrow\nn \\
 & = -\frac {1}{N_c} \beta^{DD}_{sbdd}\, O_2^d[R,R] +  \beta^{DD}_{sbdd}\, O_1^d[R,R]
     -\frac {1}{N_c} \beta^{DD}_{sddb}\, \widehat{O}_2^d[R,R] + \beta^{DD}_{sddb}\, \widehat{O}_1^d[R,R]\\
 & \phantom{=}   
    -\frac {1}{N_c} \beta^{DD}_{sbbb}\, O_2^b[R,R] + \beta^{DD}_{sbbb}\, O_1^b[R,R]
    -\frac {1}{N_c} \beta^{DD}_{sbss}\, O_2^s[R,R] + \beta^{DD}_{sbss}\, O_1^s[R,R]
\end{align}
gives 
\begin{align}
\mathcal{V} \Delta C^{s}_{1}[R,R](\mu_{KK})=\frac{1}{ T^2} \beta^{DD}_{sbss} &&
\mathcal{V} \Delta C^{s}_{2}[R,R](\mu_{KK})=-\frac{1}{N_c T^2} \beta^{DD}_{sbss}\nn \\
\mathcal{V} \Delta C^{b}_{1}[R,R](\mu_{KK})=\frac{1}{ T^2} \beta^{DD}_{sbbb} &&
\mathcal{V} \Delta C^{b}_{2}[R,R](\mu_{KK})=-\frac{1}{N_c T^2} \beta^{DD}_{sbbb}\nn \\
\mathcal{V} \Delta C^{d}_{1}[R,R](\mu_{KK})=\frac{1}{T^2} \beta^{DD}_{sbdd} &&
\mathcal{V} \Delta C^{d}_{2}[R,R](\mu_{KK})=-\frac{1}{N_c T^2} \beta^{DD}_{sbdd}\nn \\
\mathcal{V} \Delta \widehat{C}^{d}_{1}[R,R](\mu_{KK})=\frac{1}{T^2} \beta^{DD}_{sddb} &&
\mathcal{V} \Delta \widehat{C}^{d}_{2}[R,R](\mu_{KK})=-\frac{1}{N_c T^2} \beta^{DD}_{sddb}
\end{align}

\begin{align}
  b^{QQ}_{ij}  \bar Q_i& \gamma^\mu T^A Q_i  \bar Q_j \gamma_\mu T^A Q_j \longrightarrow\nn \\
 & = -\frac {1}{N_c} \beta^{QQ}_{sbuu}\, O_2^u[L,L] +  \beta^{QQ}_{sbuu}\, O_1^u[L,L]\nn \\
 & \phantom{=}   
    -\frac {1}{N_c} \beta^{QQ}_{sbdd}\, O_2^d[L,L] + \beta^{QQ}_{sbbb}\, O_1^d[L,L]
    -\frac {1}{N_c} \beta^{QQ}_{sddb}\, \widehat{O}_2^s[L,L] + \beta^{QQ}_{sddb}\, \widehat{O}_1^s[L,L]\nn \\
  & \phantom{=}  
  -\frac {1}{N_c} \beta^{QQ}_{sbss}\, O_2^d[L,L] + \beta^{QQ}_{sbss}\, O_1^s[L,L]
  -\frac {1}{N_c} \beta^{QQ}_{sbbb}\, O_2^b[L,L] + \beta^{QQ}_{sbbb}\, O_1^b[L,L]
\end{align}
gives
\begin{align}
\mathcal{V} \Delta C^{s}_{1}[L,L](\mu_{KK})=\frac{1}{ T^2} \beta^{QQ}_{sbss} &&
\mathcal{V} \Delta C^{s}_{2}[L,L](\mu_{KK})=-\frac{1}{N_c T^2} \beta^{QQ}_{sbss}\nn \\
\mathcal{V} \Delta C^{b}_{1}[L,L](\mu_{KK})=\frac{1}{ T^2} \beta^{QQ}_{sbbb} &&
\mathcal{V} \Delta C^{b}_{2}[L,L](\mu_{KK})=-\frac{1}{N_c T^2} \beta^{QQ}_{sbbb}\nn \\
\mathcal{V} \Delta C^{d}_{1}[L,L](\mu_{KK})=\frac{1}{ T^2} \beta^{QQ}_{sbdd} &&
\mathcal{V} \Delta C^{d}_{2}[L,L](\mu_{KK})=-\frac{1}{N_c T^2} \beta^{QQ}_{sbdd}\nn \\
\mathcal{V} \Delta \widehat{C}^{d}_{1}[L,L](\mu_{KK})=\frac{1}{T^2} \beta^{QQ}_{sddb} &&
\mathcal{V} \Delta \widehat{C}^{d}_{2}[L,L](\mu_{KK})=-\frac{1}{N_c T^2} \beta^{QQ}_{sddb}\nn\\
\mathcal{V} \Delta {C}^{q_u}_{1}[L,L](\mu_{KK})=\frac{1}{T^2} \beta^{QQ}_{sdq_uq_u} &&
\mathcal{V} \Delta {C}^{q_u}_{2}[L,L](\mu_{KK})=-\frac{1}{N_c T^2} \beta^{QQ}_{sbq_uq_u}
\end{align}

\noindent
Finally
\begin{align}
  b^{QD}_{ij}  \bar Q_i& \gamma^\mu T^A Q_i  \bar D_j \gamma_\mu T^A D_j \longrightarrow
  \nn \\
 & = -\frac {1}{2N_c} \beta^{QD}_{uusb}\, O_2^u[R,L] + \frac12 \beta^{QD}_{uusb}\, O_1^u[R,L]\nn \\
 & \phantom{=}   
      -\frac {1}{2N_c} \beta^{QD}_{sbdd}\, O_2^d[L,R] + \frac12 \beta^{QD}_{sbdd}\, O_1^d[L,R]
      -\frac {1}{2N_c} \beta^{QD}_{ddsb}\, O_2^d[R,L] + \frac12 \beta^{QD}_{ddsb}\, O_1^d[R,L]
   \nn \\
    & \phantom{=}  
      -\frac {1}{2N_c} \beta^{QD}_{sddb}\, \widehat{O}_2^d[L,R] + \frac12 \beta^{QD}_{sddb}\, \widehat{O}_1^d[L,R]
      -\frac {1}{2N_c} \beta^{QD}_{dbsd}\, \widehat{O}_2^d[R,L] + \frac12 \beta^{QD}_{dbsd}\, \widehat{O}_1^d[R,L]
   \nn \\
    & \phantom{=}  
       -\frac {1}{2N_c} \beta^{QD}_{sbss}\,  {O}_2^s[L,R] + \frac12 \beta^{QD}_{sbss}\, {O}_1^s[L,R]
       -\frac {1}{2N_c} \beta^{QD}_{sssb}\,  {O}_2^s[R,L] + \frac12 \beta^{QD}_{sssb}\, {O}_1^s[R,L]
   \nn \\
    & \phantom{=}  
        -\frac {1}{2N_c} \beta^{QD}_{sbbb}\,  {O}_2^b[L,R] + \frac12 \beta^{QD}_{sbbb}\, {O}_1^b[L,R]
       -\frac {1}{2N_c} \beta^{QD}_{bbsb}\,  {O}_2^bs[R,L] + \frac12 \beta^{QD}_{bbsb}\, {O}_1^b[R,L]
\end{align}
gives 
\begin{align}
\mathcal{V} \Delta C^{q_u}_{1}[R,L](\mu_{KK})=\frac{1}{2 T^2} \beta^{QD}_{q_uq_usb} &&
\mathcal{V} \Delta C^{q_u}_{2}[R,L](\mu_{KK})=-\frac{1}{2 N_c T^2} \beta^{QD}_{q_uq_usb}\nn \\
\mathcal{V} \Delta C^{d}_{1}[R,L](\mu_{KK})=\frac{1}{2 T^2} \beta^{QD}_{ddsb} &&
\mathcal{V} \Delta C^{d}_{2}[R,L](\mu_{KK})=-\frac{1}{2 N_c T^2} \beta^{QD}_{ddsb}\nn \\
\mathcal{V} \Delta C^{d}_{1}[L,R](\mu_{KK})=\frac{1}{2 T^2} \beta^{QD}_{sbdd} &&
\mathcal{V} \Delta C^{d}_{2}[L,R](\mu_{KK})=-\frac{1}{2 N_c T^2} \beta^{QD}_{sbdd}\nn \\
\mathcal{V} \Delta \widehat{C}^{d}_{1}[R,L](\mu_{KK})=\frac{1}{2 T^2} \beta^{QD}_{dbsd} &&
\mathcal{V} \Delta \widehat{C}^{d}_{2}[R,L](\mu_{KK})=-\frac{1}{2 N_c T^2} \beta^{QD}_{dbsd}\nn \\
\mathcal{V} \Delta \widehat{C}^{d}_{1}[L,R](\mu_{KK})=\frac{1}{2 T^2} \beta^{QD}_{sddb} &&
\mathcal{V} \Delta \widehat{C}^{d}_{2}[L,R](\mu_{KK})=-\frac{1}{2 N_c T^2} \beta^{QD}_{sddb}\nn \\
\mathcal{V} \Delta C^{s}_{1}[L,R](\mu_{KK})=\frac{1}{2 T^2} \beta^{QD}_{sbss} &&
\mathcal{V} \Delta C^{s}_{2}[L,R](\mu_{KK})=-\frac{1}{2 N_c T^2} \beta^{QD}_{sbss}\nn \\
\mathcal{V} \Delta C^{b}_{1}[L,R](\mu_{KK})=\frac{1}{2 T^2} \beta^{QD}_{sbbb} &&
\mathcal{V} \Delta C^{b}_{2}[L,R](\mu_{KK})=-\frac{1}{2 N_c T^2} \beta^{QD}_{sbbb}\nn \\
\mathcal{V} \Delta C^{s}_{1}[R,L](\mu_{KK})=\frac{1}{2 T^2} \beta^{QD}_{sssb} &&
\mathcal{V} \Delta C^{s}_{2}[R,L](\mu_{KK})=-\frac{1}{2 N_c T^2} \beta^{QD}_{sssb}\nn \\
\mathcal{V} \Delta C^{b}_{1}[R,L](\mu_{KK})=\frac{1}{2 T^2} \beta^{QD}_{bbsb} &&
\mathcal{V} \Delta C^{b}_{2}[R,L](\mu_{KK})=-\frac{1}{2 N_c T^2} \beta^{QD}_{bbsb}\nn \\
\end{align}


\end{document}